\documentclass[aps,prl,twocolumn,showpacs,groupedaddress,amsmath,amssymb]{revtex4}
\usepackage{graphicx}% Include figure files
\usepackage{color}
\usepackage{bm}
\usepackage{doi}
\usepackage{float}

\begin{document}
%Eigenstates Transition Without Undergoing an Adiabatic Process; 
\title{Eigenstates Transition Without Undergoing an Adiabatic Process}
\author{Fatemeh Mostafavi$^1$, Luqi Yuan$^2$, Hamidreza Ramezani$^1$} 
\email {hamidreza.ramezani@utrgv.edu}
\affiliation{$^1$ Department of Physics and Astronomy, University of Texas Rio Grande Valley, Brownsville, TX 78520, USA \\
$^2$ School of Physics and Astronomy, Shanghai Jiao Tong University, Shanghai 200240, China }

\begin{abstract}
We introduce a class of non-Hermitian Hamiltonians that offers a dynamical approach to shortcut to adiabaticity (DASA). In particular, in our proposed $2\times 2$ Hamiltonians, one eigenvalue is absolutely real and the other one is complex. This specific form of the eigenvalues helps us to exponentially decay the population in an undesired eigenfunction or amplify the population in the desired state while keeping the probability amplitude in the other eigenfunction conserved. This provides us with a powerful method to have a diabatic process with the same outcome as its corresponding adiabatic process. In contrast to standard shortcuts to adiabaticity, our Hamiltonians have a much simpler form with a lower thermodynamic cost. Furthermore, we show that DASA can be extended to higher dimensions using the parameters associated with our $2\times 2$ Hamiltonians. Our proposed Hamiltonians not only have application in DASA but also can be used for tunable mode selection and filtering in acoustics, electronics, and optics.
\end{abstract}

%\pacs{31.50.Gh,}

\maketitle

The current transition of technological advancements from classical to quantum systems, makes the quantum adiabatic theorem an important matter beyond a conceptual curiosity with widespread applications in atomic and molecular physics \cite{1,1p,2,3,4,5}, quantum Hall physics \cite{6,7}, the physics of geometric phase \cite{8}, quantum computation \cite{9,10,11}, quantum annealing \cite{12,13,14}, and quantum simulations \cite{15}. The adiabatic theorem in its earliest form \cite{16} states that a quantum system with a time-dependent Hamiltonian $H(\epsilon t)$ and non-degenerate discrete states will remain in its instantaneous ground state (GS) if it is initially prepared in its GS and its Hamiltonian changes sufficiently slow in time, namely $\epsilon \to 0$. Apart from some inconsistency for certain Hamiltonians \cite{17,18,19}, while there is no doubt about the correctness of adiabatic theorem, in practice it is very difficult, if not impossible, to satisfy its necessary conditions due to the competition between the scan time and decoherence time resulted from the existence and unavoidable undesired non-adiabatic channels. To overcome this problem and improve the population transfer from the GS of the original system to the GS of the final system \cite{note} without disturbing other states some techniques have been proposed including nonlinear level crossing \cite{20}, amplitude-modulated and composite pulses \cite{21,22}, and parallel adiabatic passage \cite{23}. Another growing approach is the so-called ``shortcuts to adiabaticity'' where one looks for fast processes with the same outcome as an ideal and yet infinitely slow process. The common approach in the shortcuts to adiabaticity is to nullify the non-adiabatic coupling by introducing the so-called counter-diabatic extra field \cite{24,25,26,27,28}. The shortcut to adiabaticity originally studied in Hermitian systems and has been extended to non-Hermitian systems \cite{29,30,31}. The rapid adiabatic passage in the above methods comes with a fundamental problem, namely the cost of increasing the coupling in the Hermitian case or adding more gain/loss rate in the non-Hermitian case and raise the question of trade-off between the speed and energy consumption (thermodynamic cost) of such methods to realize a quantum process\cite{32,33}. Furthermore, the functional form of the external parameter, including the gain and loss profile, might not be a simple function and thus it would be an extremely challenging task to create such complex functions. Therefore, it would be exceedingly important to bypass the fundamental limits and have a fast population transfer from GS to GS without disturbing other states with lower thermodynamic cost and with much simpler functions (hopefully constant!) that are feasible and experimentally accessible.

To address the above demand, in this Letter, by introducing a new class of Hamiltonians we propose a totally different approach from previous works for complete population transfer from an eigenstate (GS) of the initial system to the eigenstate (GS) of the final system without disturbing other states in an almost instantaneous manner.  In our approach which we call it dynamical approach to shortcut to adiabaticity (DASA), we focus on engineering the Hamiltonian and the dynamical properties of the system to remove (implant) any undesired (desired) probability amplitude rather than controlling the adiabatic passage and enforcing the transition to occur in an exclusive manner. %In our proposal, the projection of the starting state to the corresponding final state indicates the amount of the probability amplitude that is initially projected to the final state.
 In particular, starting from two-state quantum systems and using the method of non-Hermitian diagonalization transformation we find the value of complex part of a general $2\times 2$ Hamiltonian such that the undesired (desired) amplitude dissipates (amplifies) dynamically in an exponential manner. %For a system with special constraints, we provide a simple form for the complex parts of the Hamiltonian and discuss the field evolution in such Hamiltonians. 
Furthermore, we extend our approach to higher dimensions and we show that for a three-state Hamiltonian the same complex part as the two-state Hamiltonian results in complete population transfer in no time. Specifically, we apply our approach to the two- and three-level Landau-Zener (LZ) model and show that after an exponential transient time the system undergoes a complete population transfer, namely, it has a population at the designated state and the other state becomes empty. The amplitude and phase of the probability in the designated state depends only on the inner product of the initial state and final state. In contrast to other non-Hermitian shortcuts to adiabaticity where the non-Hermitian function is a complicated function, the complex part of our Hamiltonian is just a constant and can be implemented with different degrees of non-Hermiticity, making it easy to realize our proposal experimentally. We would like to mention that although our discussion is used to introduce the DASA, our Hamiltonians can be used for dynamical and tunable mode selection and filtering in a wide range of systems from acoustics, to electronics, to optics and photonics. 

To demonstrate DASA let us consider a two-level quantum system with a general $2\times 2$ time-independent non-Hermitian Hamiltonian of the following form
\begin{equation}
H=\sigma_x+\frac{i \Delta\gamma+\Delta\omega}{2}\sigma_z+\frac{i \Sigma\gamma+\Sigma \omega}{2} {\bf 1}
\label{eq1}
\end{equation}   
where $\Delta\gamma=\gamma_1- \gamma_2$, $\Sigma\gamma=\gamma_1+ \gamma_2$, $\Sigma \omega=\omega_1+\omega_2$, $\Delta\omega=\omega_1-\omega_2$, $\sigma_{x,z}$ are Pauli matrices, and $\mathbf{1}$ is the identity matrix. Notice that we normalized the on-site potentials $\omega_{1,2}+i\gamma_{1,2}$ to the coupling between the states. The Hamiltonian $H$ in Eq.(\ref{eq1}) has been used to model, for instance, light propagation in coupled waveguides and resonators with gain and loss \cite{salamo,lan} and dynamics of open quantum systems\cite{PTamo,antipt}. The eigenvalues of the Hamiltonian $H$ in Eq.(\ref{eq1}), here denoted by $\lambda_{1,2}$ with the corresponding eigenvectors $|\lambda_{1,2}\rangle$, are generally complex. Depending on the sign of the imaginary part of the eigenvalues, the associated eigenvector will undergo amplification or absorption. %For instance, in parity-time symmetric systems $\omega_1=\omega_2$ and $\gamma_1=-\gamma_2$ and for $|\gamma_1|> 1$ the two eigenmodes are complex, one exponentially amplifying and the other exponentially decaying. 
\begin{figure}
		\includegraphics[width=1\linewidth, angle=0]{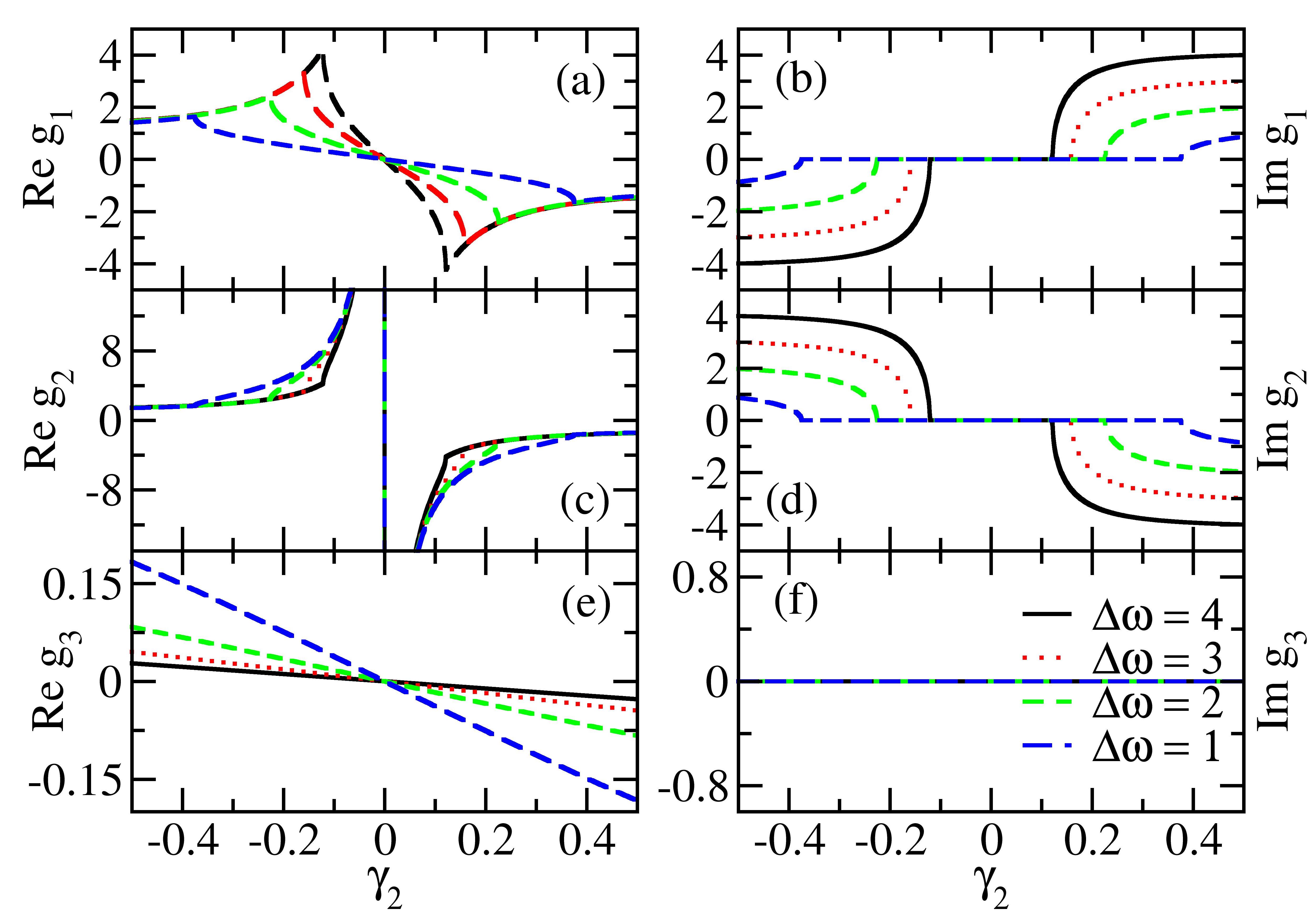}
	\caption{(Color online) Real (left column) and imaginary (right column) part of the $g_{1,2,3}$ as a function of the $\gamma_2$ for $\Delta\omega=1,2,3,4$. As long as the imaginary part is zero the value of $g$ function is acceptable, namely by choosing a $\gamma_2$ and the corresponding $\gamma_1$ from any of the functions $g_{1,2,3}$ one eigenvalues of Hamiltonian in Eq.(\ref{eq1}) becomes absolutely real and the other one becomes complex. Notice that the three functions have different behavior. For example, $g_3$ is always real, or $g_2$ becomes very large around $\gamma_2=0$. \label{fig1}}
\end{figure}

Any initial excitation $|\psi (0)\rangle$ can be written as a superposition of $|\lambda_{1,2}\rangle$, namely $|\psi (0)\rangle=c_1|\lambda_{1}\rangle+c_2|\lambda_{2}\rangle$ and evolves in time according to  
\begin{equation}
|\psi (t)\rangle=e^{-iHt}|\psi (0)\rangle= c_1e^{-i\lambda_1 t}|\lambda_{1}\rangle+c_2 e^{-i\lambda_2 t}|\lambda_{2}\rangle.
\label{eq2}
\end{equation}
Here we are interested to find the eigenmodes such that one eigenmode decays or amplifies while the other mode remains unchanged. Therefore, in Eq.(\ref{eq2}) one eigenvalue should be completely real while the other one is complex. Imposing such a constraint to the Hamiltonian in Eq.(\ref{eq1}), we expect its eigenvalues, associated with the entries of the diagonal matrix $H_d$, have the following form:
\begin{equation}
\lambda_1=x_1+i y,\quad \lambda_2=x_2
\label{eq3}
\end{equation}
where $x_{1,2}$ and $y$ are some real parameters that we aim to find them. Specifically, following the eigen decomposition identity the square matrix $H$ in Eq.(\ref{eq1}) can be decomposed into the very special form 
\begin{equation}
H=R^{-1} H_d R,\quad R\equiv\left(
\begin{array}{cc}
a&b\\
c&d
\end{array}
\right )
\label{eq4}
\end{equation}
where $R$ is a matrix composed of the eigenvectors of $H$ and $R^{-1}$ is the inverse matrix of $R$\cite{note2}. In order to find the $\lambda_{1,2}$ and the corresponding $|\lambda_{1,2}\rangle$ eigenvectors, we assume $a,b,c,d$ are free and unknown parameters. From similarity transformation in Eq.(\ref{eq4}), it is easy to show that $a=-\frac{bd}{c}$. By replacing $a$ in Eq.(\ref{eq4}) and solving for parameter $d$ we get two solutions of the following form 
$
d=\pm\frac{c \sqrt{-i \gamma_1+x_2-\omega_1}}{\sqrt{i \gamma_1-x_1-i y+\omega_1}}$ or  $d=\pm\frac{c \sqrt{x_1+i (-\gamma_2+y+i \omega_2)}}{\sqrt{i \gamma_2-x_2+\omega_2}}$. As $d$ is a unique parameter by equating these solutions and solving for $\gamma_1$, we get $\gamma_1=-\gamma_2+i (\Sigma \omega-x_1-x_2)+y$. However, $\gamma_1$ is a real parameter, thus $\Sigma \omega-x_1-x_2=0$. Therefore we come to the conclusion that
\begin{equation}
x_1+x_2=\Sigma \omega,\quad y=\Sigma \gamma.
\label{eq5}
\end{equation} 
Equation (\ref{eq5}) recovers the well-known fact that the trace of a matrix is invariant under the similarity transformation in Eq.(\ref{eq4}). Using Eq.(\ref{eq5}) together with the off-diagonal terms in Eq.(\ref{eq4}) and the assumption that the coupling is real we can easily show
\begin{equation}
x_1=\frac{\omega_1 \gamma_1+\omega_2 \gamma_2}{\gamma_1+\gamma_2},\quad x_2=\frac{\omega_1 \gamma_2+\omega_2 \gamma_1}{\gamma_1+\gamma_2}.
\label{eq6}
\end{equation}
From Eq.(\ref{eq6}) it is clear that the real part of the eigenvalues can be exchanged by replacing $\omega_1\leftrightarrow\omega_2$ or $\gamma_1\leftrightarrow\gamma_2$. We can plug the solutions given by Eq.(\ref{eq6}) into the Eq.(\ref{eq4}) and look for $b$ which leads to $b= \frac{-i a y}{\gamma _1 (y-i \Delta\omega)}$ or $b=\frac{a \gamma_2 (iy-\Delta\omega)}{y}$. By equating these solutions we can find three solutions $g_{1,2,3}$ \cite{Supp} for $\gamma_1$ as a function of $\Delta\omega$ and $\gamma_2$.
The parameter $\Delta \omega$ in the $g_{1,2,3}$ always appears with a square power. Therefore the value of $\gamma_1$ is invariant under the change in the sign of the $\Delta\omega$. Furthermore, because all the solutions in $g_{1,2,3}$ have a part proportional to $1/\Delta\omega$, there is no Hamiltonian with the above properties when $\Delta\omega=0$. Thus, our system cannot be mapped to a parity-time symmetric one. We have plotted in Fig.(\ref{fig1}) the real and imaginary parts of the $g_{1,2,3}$ functions for several values of $\Delta \omega$ as a function of $\gamma_2$. While the function $g_3$ is always real, the other two functions, $g_{1,2}$, might be complex depending on the value of $\Delta \omega$ and $\gamma_2$. However, originally we assumed that $\gamma_1$ is real, therefore, for functions $g_{1,2}$ we should confine ourselves to the domains that $g_{1,2}$ are real. From these solution we observe that the Hamiltonian $H$ in Eq.(\ref{eq1}) with eigenvalues of the following forms \begin{equation}
\lambda_1=\frac{\omega_1\gamma_1+\omega_2\gamma_2}{\gamma_1+\gamma_2}+i(\gamma_1+\gamma_2),\quad \lambda_2=\frac{\omega_1\gamma_2+\omega_2\gamma_1}{\gamma_1+\gamma_2}
\label{eq8}
\end{equation} 
is not unique and can be built using any of the $g_{1,2,3}$ functions. Clearly, for any choice of the $g$ function the system has different eigen energies and amplification or dissipation. The amplification or dissipation defines by the sign of the $\gamma_1+\gamma_2$. Specifically, from Eq.(\ref{eq2}) we infer that during the evolution of the original wave-packet, in the wave-function $|\psi(t)\rangle$ the part that is proportional to the eigenstate $|\lambda_1\rangle$ undergoes an exponential decay (amplification) if $\gamma_1+\gamma_2<0(>0)$ while the portion associated with $|\lambda_2\rangle$ remains conserved and only accumulates a phase. 

%The functions $g_{1,2,3}$ has a complicated form, however, if we impose the two constraints, namely $\Delta\gamma^2-\Delta\omega^2=4$ and $\gamma_1+\gamma_2=\sqrt{\Delta\omega \Delta\gamma}$ where $\Delta\gamma=\gamma_1-\gamma_2$, we can easily show that the complex part of the Hamiltonian $H$ in Eq.(\ref{eq1}) will follow a more simpler form, namely, $\gamma_1=\frac{1}{2} (\pm\sqrt{\Delta\omega^2+4}\mp\sqrt{\Delta\omega\sqrt{\Delta\omega^2+4} })$ and $\gamma_2=\frac{1}{2} (\mp\sqrt{\Delta \omega ^2+4}\mp\sqrt{\Delta \omega  \sqrt{\Delta \omega ^2+4}})$ and the eigenvalues become $\lambda_1=\frac{\Sigma \omega+\Sigma\gamma}{2}+i\Sigma\gamma$ and $\lambda_2=\frac{\Sigma \omega-\Sigma\gamma}{2}$. Notice that irrespective of the complex form of the $g_{1,2,3}$ functions or the above-simplified forms under specific conditions, in our Hamiltonian the values of all parameters including $\gamma_{1,2}$ are simply some constants.

Before discussing the properties of the eigenstates of the Hamiltonian $H$ we would like to note that by a proper choice of $\omega_{1,2}$ and $\gamma_{1,2}$ in Eq.(\ref{eq8}) one can make the lower (higher) energy to be complex and higher (lower) energy to be real. This property of our proposed Hamiltonians helps us to have population transfer from any state to upper or lower state only by a correct choice of the mentioned parameters. Now that we know the exact form of the Hamiltonian $H$ in Eq.(\ref{eq1}) and the corresponding eigenvalues in Eq.(\ref{eq8}), we can easily calculate the eigenstates $|\lambda_{1,2}\rangle$
\begin{equation}
|\lambda_1\rangle=\left(\begin{array}{c}
\frac{y}{\gamma_2(iy-\Delta\omega)}\\1
\end{array}\right), \quad |\lambda_2\rangle=\left(\begin{array}{c}
1\\\frac{-\gamma_1(\Delta\omega+iy)}{y}
\end{array}\right).
\label{eq9}
\end{equation}

Armed with the exact form of the eigenvectors and eigenvalues of our proposed class of non-Hermitian Hamiltonian, we can discuss the dynamics associated with our system and show how one can achieve DASA in less than no time and with lower cost by a correct choice of our simple proposed Hamiltonians. Specifically, we are interested in the problem of population transfer from bare state $|1\rangle=(0,1)^T$ to $|2\rangle=(1,0)^T$ given by well-celebrated LZ model with constant coupling, namely 
$
H_{LZ}(\epsilon,t)=\sigma_x-(\epsilon^2 t)\sigma_z$
where $t$ is time which spans from $-\infty$ to $\infty$ and $\epsilon$ is a positive real parameter that determines the adiabaticity of the process. Namely, from adiabatic theorem we expect that a complete population transfer from lower energy state $|1\rangle=(0,1)^T$ at $t=-\infty$ to the GS $|2\rangle=(1,0)^T$ at $t=\infty$ occurs for small values of $\epsilon$. For larger values of $\epsilon$ the population transfer to the state $|2\rangle=(1,0)^T$ at $t=\infty$ becomes smaller\cite{Supp}.% In Fig.(\ref{fig2}a,b) we plotted the population of the bare states for different values of $\epsilon$ in time interval $[-15,15]$ (in the unit of coupling) which clearly confirms our expectation. 
\begin{figure}
	\includegraphics[width=1\linewidth, angle=0]{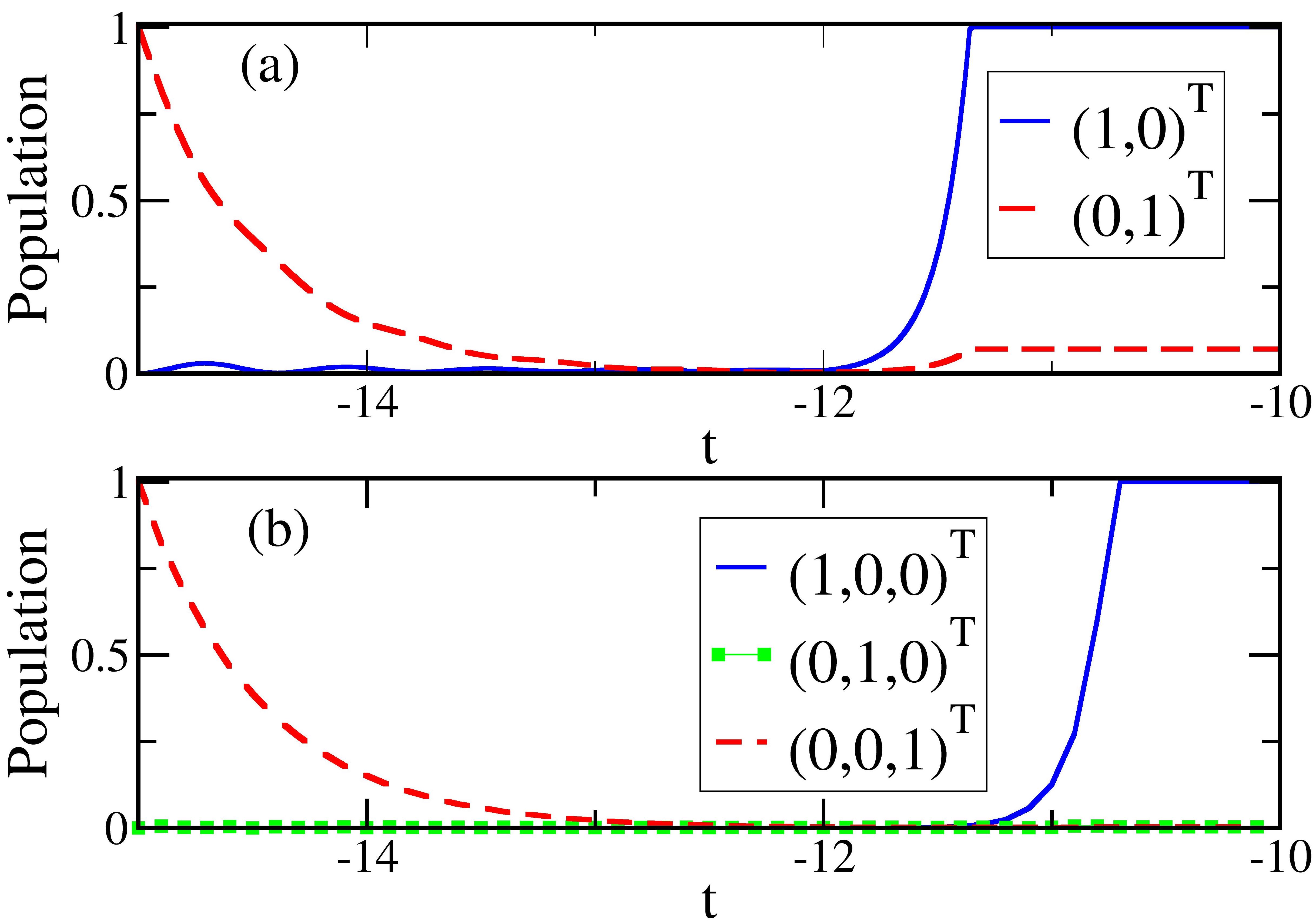}
	\caption{(Color online) (a) Amplitude probability of the bare states $(0,1)^T$ and $(1,0)^T$ as a function of time (in the unit of coupling) using the Hamiltonian ${\cal H}_{2\times 2} (t)$. (b) Amplitude probability of the bare states $(0,0,1)^T$, $(0,1,0)^T$, and $(1,0,0)^T$ as a function of time using the Hamiltonian ${\cal H}_{3\times 3} (t)$. In both (a,b) the same gain and loss parameters are incorporated. Furthermore, in both (a,b) the initial excited state is the GS while at the end of the process the populated state is the GS.  \label{fig2}}
\end{figure}

Now let's see how one can get complete population transfer from GS of the old system to GS of the new system without disturbing other states in a very short time via our proposed class of Hamiltonians. As discussed earlier depending on the values of $\Sigma\gamma$ being a positive or negative constant, probability amplitude in one eigenstate will undergo dissipation or amplification. Therefore, if we are able to find parameters of Hamiltonian $H$ in Eq.(\ref{eq1}) such that its eigenvector with lower energy becomes $|\lambda_1\rangle\approx|1\rangle=(0,1)^T$ and $\Sigma\gamma<0$, then we expect an exponential decay at state $|1\rangle$. On the other hand, the original population in the other bare state, namely $|\lambda_2\rangle\approx|2\rangle=(1,0)^T$, will remain constant. Nevertheless, as the two states $|1\rangle$ and $|2\rangle$ are orthogonal, the constant population in the bare state $|2\rangle$ is very small. To amplify the amplitude in the bare state $|2\rangle$ we can introduce another Hamiltonian where this time the parameters are chosen such that its eigenvector with lower energy becomes approximately equal to $|2\rangle=(1,0)^T$ and $\Sigma\gamma>0$ which results in an exponential amplification at state $|2\rangle$. We can cut the second Hamiltonian after probability amplitude in $|2\rangle$ becomes one. 

To find the correct parameters lets go back to Eq.(\ref{eq8}) and Eq.(\ref{eq9}) where we have the exact form of the eigenvalues and eigenvectors of the Hamiltonian $H$. For $\omega_1=0$ from Eq.(\ref{eq8}) we find that if $|\gamma_2|>|\gamma_1|$, $\gamma_{2(1)}<(>)0$, and $\omega_2<0$ then $|\lambda_1\rangle$ is the lower energy level. One can show that if $|\gamma_1+\gamma_2|\ll |\omega_2|$ then $|\lambda_1\rangle\approx(0,1)^T=|1\rangle$. Similarly, using the same equations one can show that for $\omega_2=0$, $\omega_1\approx0$, and $\gamma_1>|\gamma_2|\approx0$, where both $\omega_1,\gamma_2<0$, the eigenstate $|\lambda_1\rangle\approx|2\rangle=(1,0)^T$ becomes the lower energy level and undergoes an exponential amplification.
An example of such process is depicted in Fig.(\ref{fig2}a) where initially (at $t=-15$) we exited the bare state $|1\rangle$ and solved the Schr\"{o}dinger equation with Hamiltonian 
%\begin{equation}
%\begin{array}{c}
${\cal H}_{2\times 2}(t)=H_1\times (\mathbf{\Theta}(t+15)-\mathbf{\Theta}(t+12))+
H_2\times (\mathbf{\Theta}(t+12)- \mathbf{\Theta}(t+11.358))+
\mathbf{1}\times \mathbf{\Theta}(t+11.358)$
%\end{array}
%\label{eqH2}
%\end{equation}
Where $\mathbf{\Theta}(x)=
\begin{cases}
\mathbf{0}       & \quad \text{if } x<0 \\
\mathbf{1}  & \quad \text{if } x >0
\end{cases}$ is the Heaviside step function matrix, with $\mathbf{0}$ as the zero matrix. The two matrices $H_{1,2}$ have the form of Hamiltonian in Eq.(\ref{eq1}) where the parameters are chosen to be $\omega_1+i\gamma_1=  0+ig_3(10,-0.95)$, $\omega_2+i\gamma_2=-10-0.95i$ and in $H_2$ we chose $\omega_1+i\gamma_1= - 0.01+ig_2(-0.01,-0.25)$, $\omega_2+i\gamma_2=0-0.25i$. 
%\begin{equation}
%	H_1=\left(\begin{array}{cc}
%	0+ig_3(10,-0.95)&1\\
%	1&-10-0.95i
%	\end{array}\right)
%\end{equation}
%and 
%\begin{equation}
%H_2=\left(\begin{array}{cc}
%- 0.01+ig_2(-0.01,-0.25)&1\\
%1&0-0.25i
%\end{array}\right)
%\end{equation}
%$\omega_1+i\gamma_1=  0+ig_3(10,-0.95)$, $\omega_2+i\gamma_2=-10-0.95i$ and in $H_2$ we chose $\omega_1+i\gamma_1= - 0.01+ig_2(-0.01,-0.25)$, $\omega_2+i\gamma_2=0-0.25i$. 
We observe that a complete population transfer (from GS to a new GS) occurs after two consequent exponential transition times, wherein one the probability amplitude in $(0,1)^T$ (at the lower energy) decays and in the second one, state $(1,0)^T$ (with the lower energy) amplifies. It should be mentioned that by including more loss (gain) in the first (second) transition time, one can shorten the transition times even more. This is the price that one pays for faster transitions similar to the other non-Hermitian shortcuts to adiabaticity methods \cite{30}. However, two major differences exist, first the rate of the gain and loss in our system is significantly smaller than the previous methods, namely our method has a lower cost. For example, in order to have a complete population transfer in $6$ coupling units using the common approach one needs to incorporate a total gain equal to $\approx -5$ during the process with the maximum value of gain equal to $-12$ (which is usually very difficult to reach in reality) at $t=0$, while in our case we need to have a total gain equal to $\approx -2.6$ in $3.6$ coupling time units \cite{30,Supp} with maximum value of gain $\approx -3.99$. Thus, our process needs $\approx 2$ orders less gain for a twice faster process. Secondly, the gain and loss profiles in the mentioned methods follow a complicated form, while in our case the gain or loss are constant numbers which makes it much more experimental friendly.

A question that might arise is how one can implement our approach in higher dimensions. In extended LZ model to higher dimensions \cite{Supp} the middle levels play and important role. If the on-site potentials of middle levels become larger and larger the adiabatic process needs to get slower and slower\cite{Supp}. Interestingly enough for the same cases with large on-site potentials of the middle levels, one can show that the same gain and loss parameters that we provided for DASA in the two-level system will result in DASA in higher dimensions without any extra effort \cite{Supp,adiabatic2}. Note that although in higher dimensions we lose the special form of the eigenvalues for a two-level system (one real the other complex), the complex form of the eigenvalues come to our benefit \cite{Supp}. An example of such process is given in Fig.(\ref{fig2}) where we have solved the Schr\"{o}dinger equation for a three-level system with $
{\cal H}_{3\times 3}(t)=H_3\times (\mathbf{\Theta}(t+15)-\mathbf{\Theta}(t+12))+
H_4\times (\mathbf{\Theta}(t+12)- \mathbf{\Theta}(t+10.7374 )+
\mathbf{1}\times \mathbf{\Theta}(t+10.7374 )
$ where \begin{equation}
H_3=\left(\begin{array}{ccc}
0+ig_3(10,-0.95)&1&0\\
1&15&1\\
0&1&-10-0.95i
\end{array}\right)
\label{eq20}
\end{equation}
and 
\begin{equation}
H_4=\left(\begin{array}{ccc}
- 0.01+ig_2(-0.01,-0.25)&1&0\\
1&15&1\\
0&1&-0.25i
\end{array}\right).
\label{eq21}
\end{equation} In this example, in the first segment of the dynamics ($-15<t<-12$) the intensity in the GS of the old system decays due to the large negative imaginary part of the corresponding eigenvalue. The complex part of the other states are very small and do not result in significant amplification or absorption. In the second part of the dynamics ($-12<t<-10.7374$) the GS of the new system has a strong amplification while the other state do not have significant decay or amplification.

In conclusion, we proposed a class of $2 \times 2$ non-Hermitian Hamiltonians that have peculiar eigenvalues, one being real and the other being complex. The complex eigenvalue causes decay or amplification in the probability amplitude of the associated eigenstate while probability amplitude of the other eigenstate remains constant. The formation of such eigenvalues helps us to propose a new method for the shortest shortcut to adiabaticity which is totally different from what has been proposed so far. In contrast to the other methods associated with standard shortcuts to adiabaticity, our approach has lower cost, and is generated by a very simple Hamiltonian. Furthermore, we have shown that our $2 \times 2$ non-Hermitian Hamiltonians can be used to extend our method to higher dimensions without any further effort.

DASA can be implemented in different experimental setups such as QED\cite{QED}, coupled waveguides \cite{natPT}, acoustics \cite{alu,mine}, and electronics \cite{shindler}. For example an optical system composed of two coupled waveguides \cite{Supp} can be used to experimentally demonstrate our proposal where each waveguide has two segments. In the first segment between $z=(0,z_1)$ the gain (loss) waveguide has index of refraction $n^{g(l)}(z)=n_1^{g(l)}-(+)i\gamma_1^{g(l)}$ while in the second segment between $z=(z_1,z_2)$ it has index of refraction $n^{g(l)}(z)=n_2^{g(l)}-(+)i\gamma_2^{g(l)}$, where $z$ is the propagation direction length playing the role of $t$ in Schr\"{o}dinger equation. For example, experimentally one can obtain a coupling length as low as $1mm$ as well as a gain/loss level below $\pm30cm^{-1}$ without changing the real part of the index of refraction. For a higher value of the gain and loss, $n_{1,2}$ are affected by the imaginary part of the index of refraction through the Kramers-Kronig relation \cite{kk}. %Similarly one can observe our proposed approach using electronic circuits as described in \cite{Supp}.

%{\it Acknowledgments --} 
%We thank G. V. Naik for helpful conversations. H.R %gratefully acknowledge support from the UT system under %the Valley STAR award.

%==================================================================================


\begin{thebibliography}{99}
\bibitem{note} We do not need to confine ourselves to population transfer from GS of the old system to the GS of the final system. In general one can consider the problem for population transfer for any arbitrary state.
\bibitem{note2} The columns of the matrix $R$ are given by the $|\lambda_{1,2}$ in Eq.(\ref{eq9}). One can see easily that $\text{Det}(R)$ is not zero and thus, $R$ will not be singular.
\end{thebibliography}
\end{document}